\documentclass[preprint2]{aastex}

\shorttitle{High-resolution imaging at  SOAR}
\shortauthors{Tokovinin et al.}

\begin{document}

\title{High-resolution imaging at the SOAR telescope}

\author{A. Tokovinin, R. Cantarutti, R. Tighe, P. Schurter, N. van der
  Bliek, M. Martinez, E. Mondaca}

\affil{Cerro Tololo Inter-American Observatory, Casilla 603, La Serena, Chile}
\email{atokovinin@ctio.noao.edu}

\begin{abstract}
Bright single  and binary stars  were observed at the  4.1-m telescope
with a  fast electron-multiplication camera  in the regime  of partial
turbulence correction by the visible-light adaptive optics system.  We
compare  the  angular  resolution  achieved  by  simple  averaging  of
AO-corrected   images  (long-exposure),  selection   and  re-centering
(shift-and-add  or ``lucky'' imaging)  and speckle  interferometry. 
The  effect of  partial  AO  correction,  vibrations, and  image
post-processing on  the attained resolution is shown.   Potential usefulness of
these techniques  is evaluated for  reaching the diffraction  limit in
ground-based  optical   imaging.  Measurements of 75 binary stars  obtained
during  these tests  are given  and  objects of  special interest  are
discussed.   We   report  tentative  resolution   of  the  astrometric
companion to $\zeta$~Aqr~B. A concept of advanced high-resolution
camera is outlined.
\end{abstract}




\section{Introduction}
\label{sec:intro}

Progress of observational astronomy  depends mostly on advances in the
wavelength coverage, sensitivity, and angular resolution. Although the
record in resolution is  set by long-baseline interferometers, imaging
and  spectroscopy  of faint  sources  is  done  today at  the  angular
resolution of  adaptive optics (AO) in  the infra-red (IR)  and at the
resolution  of  the  2.4-m  Hubble  Space Telescope  in  the  visible.
Reaching  the diffraction  limit in  the visible  from the  ground for
progressively fainter  sources continues to  be an important  goal for
astronomy. Here we report on new experiments in this direction.

Diffraction-limited  resolution at  large ground-based  telescopes has
been  first attained by  speckle interferometry  \citep{Labeyrie70}. In
the  following years,  modifications  to this  method  to obtain  true
images,  rather   than  their  auto-correlations   (ACFs),  have  been
developed. One of the early approaches consisted in shifting the images
to  concidence   of  their  brightest  pixels   and  accumulating  the
result. This {\em  shift-and-add} (SAA) method \citep{Bates80,Bates82}
has  been used  to  reconstruct  the first  image  of stellar  surface
\citep{Lynds76}.   It  has been  demonstrated  that  this approach  is
superior  to   simple  tip-tilt   compensation  \citep{Christou91}.
Tip-tilt correction works well only at moderate $D/r_0$ ratios ($D$ --
telescope  diameter,   $r_0$  --  Fried  parameter);   in  this  case,
 selection of  the sharpest  images with  prominent central
speckle  is a good strategy.   For  $D/r_0  >  10$,  the  probablity  of  obtaining
diffraction-limited   image  without  turbulence   correction  becomes
negligible, but the SAA method and speckle interferometry still work.

The development of  electron-multiplication CCD (EMCCD) detectors made
it  practical to  record sequencies  of short-exposure  frames without
readout noise penalty.   On this occasion, the SAA  method was revived
under a new name,  {\em lucky imaging} \citep{Baldwin08}. A combination
of  re-centering on  bright  pixel  and selection  of  best frames  was
proposed. The selection works well only at small $D/r_0$, restricting
this approach to  long wavelengths and to telescopes  with $D < 2.5$\,m.
However, partial  turbulence correction with  AO effectively increases
$r_0$,  and  the SAA  method  can  give  good results  at  larger
apertures  \citep{Law09}. Dedicated  high-resolution  imagers such  as
LuckyCam \citep{LuckyCam} or FastCam \citep{FastCam}  were built to take
advantage of these new opportunities. 

An AO system for partial image correction in the vsible, SOAR Adaptive
Module  (SAM), is  being  built  at CTIO  \citep{SAM,  Tok08} for  the
Southern  Astrophysical Research  Telescope  (SOAR).  This  instrument
will  use a  Rayleigh  laser  guide star  to  selectively correct  low
turbulence layers. The diffraction limit  will not be reached, but the
effective increase of $r_0$ resulting from the partial correction will
boost the  potential of other  methods such as  speckle interferometry
and  SAA.  First  tests  of these  methods  at SOAR  are described  by
\citet{HRCam} and \citet{Mercury}, respectively.

In this paper  we report some results from the first  tests of SAM. At
this stage, natural guide stars (NGS)  were used to close the AO loop.
We   begin    with   the    description   of   our    experiments   in
Section~\ref{sec:exp}   and  the  analysis   of  the   AO  performance
in Section~\ref{sec:loop}.   Then three methods  of data  processing --
simple averaging,  SAA, and speckle interferometry --  are compared in
Section~\ref{sec:img}. Measurements of  close binary stars obtained as
a  by-product of  these tests  are reported  in Section~\ref{sec:res};
they  continue  the  series   of  speckle  observations  published  by
\citet{TMH10} (hereafter TMH10).  Some pairs deserve special comments;
two faint tertiary companions in  wider binaries were resolved for the
first time.  We conclude in Section~\ref{sec:concl} with a perspective
of diffraction-limited imaging in the visible.

\section{Description of the experiments}
\label{sec:exp}

SOAR is located at the Cerro Pach\'on observatory in Chile, at 2800\,m
above the  sea level.  It has  alt-azimuth mount with  two Nasmyth and
several bent Cassegrain  focal stations fed by a  flat mirror M3. This
mirror can be actuated in tip and tilt.

The  SOAR Adaptive  Module (SAM)  is an  AO system  for correcting
wave-front  distortions   and  feeding  improved   images  to  science
instruments  \citep{SAM,Tok08}.  SAM  has  internal re-imaging  optics
(two off-axis paraboloids), turbulence corrector (60-element curvature
deformable mirror,  DM), and a 10x10  Shack-Hartmann wave-front sensor
(WFS).   The SAM instrument  is designed  to work  with a  laser guide
star, but initially  it used natural guide stars  (NGS).  The light of
an on-axis star  was mostly reflected to the WFS  by a neutral beamsplitter
placed after  the DM, with  14\% transmitted to the  science detector.
The latter  was a High-Resolution Camera  (HRCAM) which contained  an EM
CCD with a  pixel scale of 15.23\,mas and  filters \citep{HRCam}. Owing
to the  $2.1^m$ light  attenuation by the  beamsplitter, we  observed only
relatively bright stars.

SAM was installed  in August 2009 at the  Nasmyth focus, receiving the
light after four  reflections in the telescope. It  rotates as a whole
for compensating field rotation as the telescope tracks stars.  The AO
loop cycle was  4.3\,ms, with a 2-cycle delay  between WFS measurement
and correction.  To compensate for  this relatively slow loop, we used
the Smith predictor  (SP) controller \citep{Madec98}. Alternatively, a
standard leaky  integrator (INT) with a  gain of 0.25  was tried.  The
degree of  AO correction is  equivalent to compansation of  45 Zernike
modes.

Of  the three engineering  runs assigned  to these  tests in  August -
October  2009, the  first was  completely lost  to clouds,  the second
produced  1.5 nights  of data  under fast  and poor  seeing of  $1'' -
1.6''$, the  third and last run  enjoyed few hours  of good conditions
(seeing 0\farcs5  to $1''$). 

We pointed telescope to single or multiple bright stars, closed the AO
loop and recorded, typically, 400  images with 20-ms exposure time and
EM  gain of  44.  Only  the  central region  of $3''  \times 3''$  was
recorded in  most cases.  In  parallel, we frequently saved  data from
the  AO loop  (sequences of  centroids and  DM voltages)  for off-line
analysis.

\section{Analysis of the AO loop data}
\label{sec:loop}

\begin{figure}[ht]
\epsscale{1.0}
\plotone{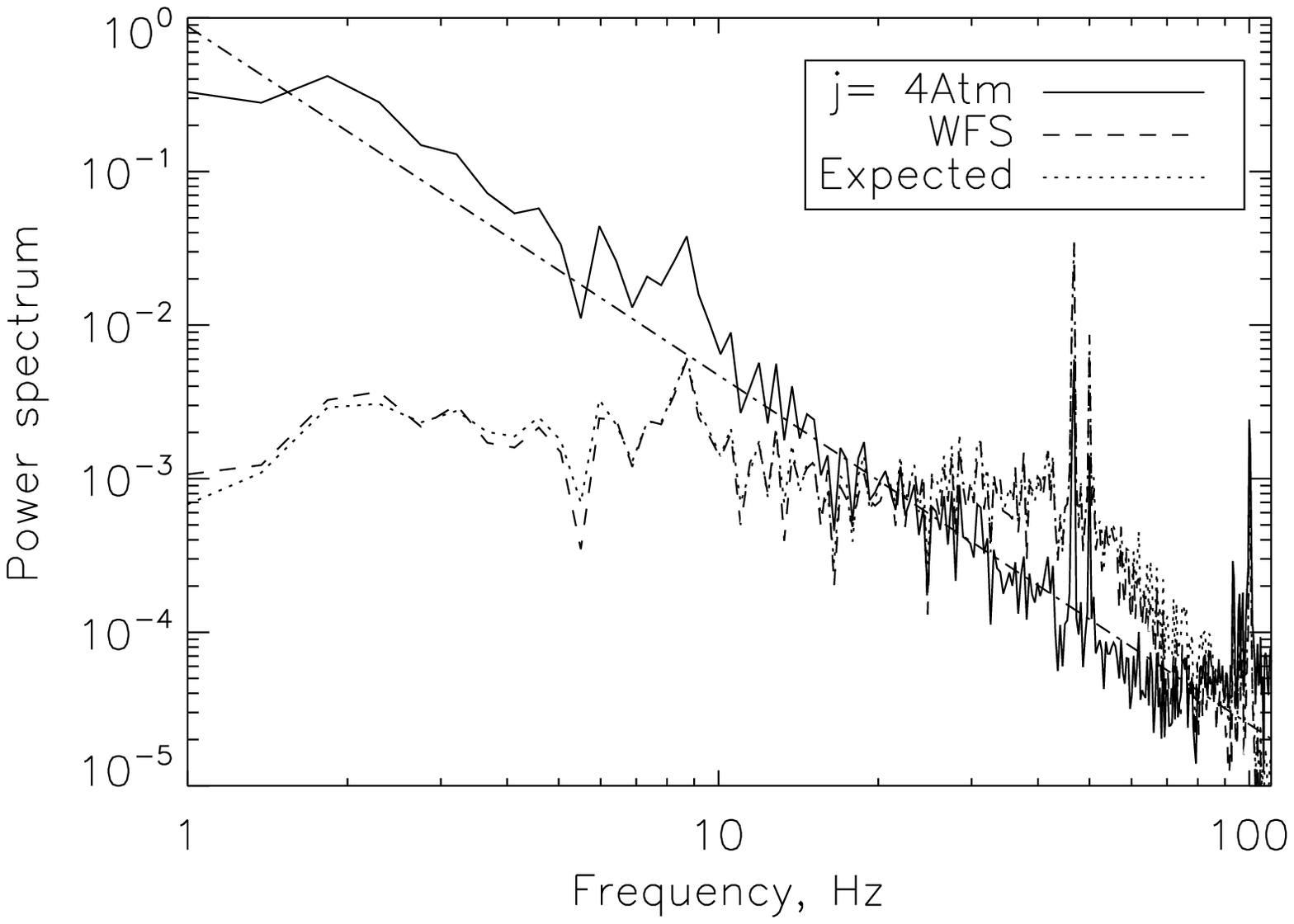} 
\plotone{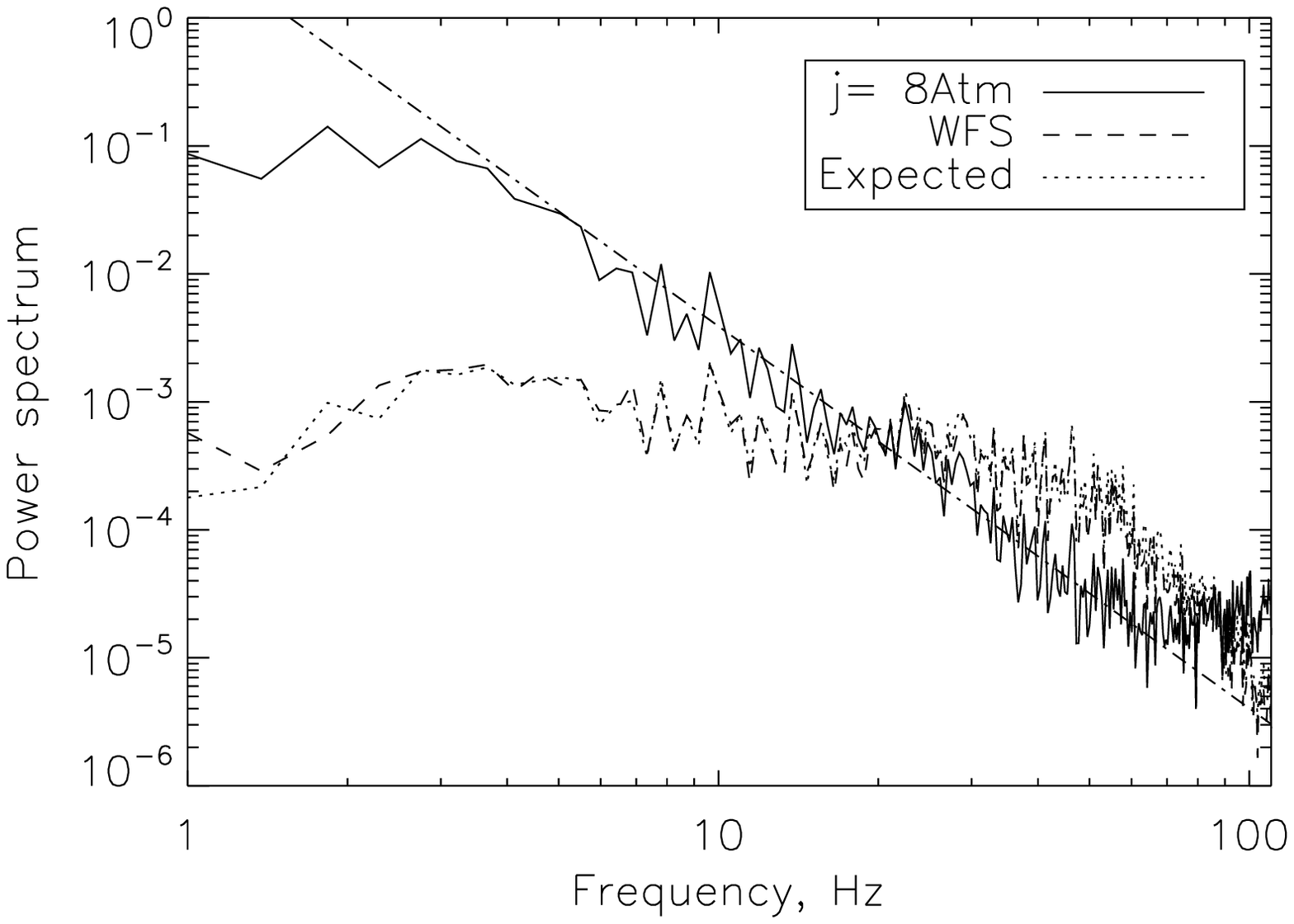} 
\caption{Temporal  power spectra of  Zernike coefficients  $a_4$ (defocus,
  top) and  $a_8$ (coma,  bottom) recorded on  October 2, 2009  under good
  seeing  ($r_0 =  0.13$\,m) in  closed loop  with the  SP controller.
  Dash-dotted straight lines indicate the power-law slopes $k=2.3$ and
  $k=3.0$, respectively.   The spectrum of  $a_4$ has lines  at 47\,Hz,
  50\,Hz, and 94\,Hz.
\label{fig:pow}  }
\end{figure}

We  calculate the  first 45  Zernike coefficients  $a_j$ of  the input
(atmospheric)  and  corrected (residual)  wave-fronts  (in radians  at
$\lambda =  0.5 \mu$m)  from the  AO loop data.   The DM  voltages are
converted  into corresponding slopes  by matrix  multiplication, using
the recorded AO interaction matrix. The slopes, in turn, are converted
into  Zernike  coefficients by  means  of  the calculated  theoretical
slopes  of Zernike  modes (gradient  matrix); we  simply  multiply the
vector of  slopes by the  pseudo-inverse gradient matrix.   Of course,
the Zernike  coefficients $a_j$ derived  in this way  contain errors  due to
noise on  the measured slopes  and cross-talk with  higher-order modes
(aliasing). 

The Fried  radius $r_0$ and seeing $\varepsilon_0  = 0.98 \lambda/r_0$
are estimated from the  variance of Zernike coefficients $\sigma^2_j =
\langle \Delta a_j^2 \rangle $ which should, in theory, equal $c_{j,j}
\; (D/r_0)^{5/3}$  \citep{Noll76}. The ratio of  measured variances to
the Noll's coefficients  $c_{j,j} $ is averaged using  modes from 7 to
19  to get  an estimate  of $D/r_0$.   The atmospheric  coherence time
$\tau_0$  is  derived  from  the  temporal variation  of  the  defocus
coefficient  $a_4$,  as  described  by \citet{FADE}.   Its  values  at
$\lambda =  0.5 \mu$m  ranged from 1\,ms  to 3.5\,ms.  Both  $r_0$ and
$\tau_0$ derived from  the loop data are in  reasonable agreement with
simultaneous measurements from the MASS-DIMM site monitor \citep{SPIE10}.

\begin{figure}[ht]
\epsscale{1.0}
\plotone{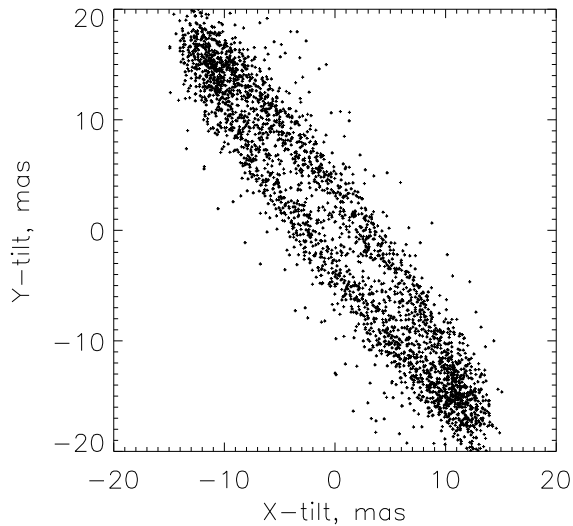} 
\caption{The 50-Hz vibration  in tip  and  tilt.  The  signal from 48\,Hz to 52\,Hz
  is isolated by the temporal bandpass filter applied
  to the $a_2$ and $a_3$ coefficients, converted to angular units, and
  plotted as XY trajectory of the optical axis.
\label{fig:tilt}  }
\end{figure}

\begin{figure*}[ht]
\epsscale{1.2}
\plotone{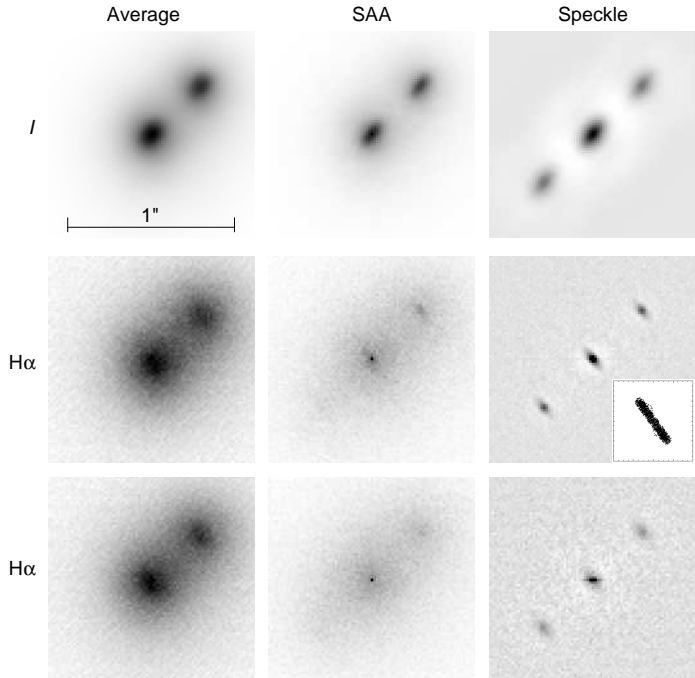} 
\caption{Comparison of  various processing algorithms for  the case of
  the  binary star  BU~172 (0\farcs408,  $V  = 6.4;  6.6$).  The  left
  column shows  average images, the  middle column -- SAA  images with
  20\%  selection, the  right column  -- ACF  obtained in  the speckle
  processing of  the same data cube.   The rows show  three data cubes
  obtained  one after  another in  a rapid  sequence. Top  row  -- $I$
  filter,  INT  controller;  middle   row  --  H$\alpha$  filter,  INT
  controller;  bottom row  --  H$\alpha$ filter,  SP controller.   The
  insert in  the middle row  shows the 50-Hz vibration  tajectory from
  the   simultaneous   loop   data,   on  a   different   scale   (cf.
  Fig.~\ref{fig:tilt}) which  matches the elongation of  the ACF peaks
  in angle  and amplitude.  The PSF  in the upper row  is elongated by
  the atmospheric dispersion.
\label{fig:BU172}  }
\end{figure*}

Temporal power  spectra of  the Zernike coefficients  corresponding to
the  input  (un-corrected)  wave-fronts  are calculated  from  the  DM
signal, with  division of  the raw power  spectrum by  the theoretical
frequency response of the closed  AO loop.  The high-frequency part of
those  spectra  shows  a  typical power-law  decline  proportional  to
$\nu^{-k}$, with exponents $k$ between 2 and 3 (Fig.~\ref{fig:pow}). A
high-frequency asymptotic decline with $k = 11/3$ is characteristic of
atmospheric  tilts,  while  a  steeper  dependence  with  $k=17/3$  is
expected for the defocus and higher modes \citep{Roddier93}. This mismatch
between  theory and  real spectra  has also been  observed  with physically
simulated turbulence in  the laboratory and  reproduced in the
numerical simulations. Most probably,  it is caused by the atmospheric
errors  of   WFS  slope  measurements  \citep{Thomas06}   and  by  the
aliasing of higher-order modes, as shown by \citet{VP10}. 

The power  spectra of the  residual coefficients derived from  the WFS
signal  show that  distortions with  low temporal  frequencies  (up to
20\,Hz) are  compensated, while high frequencies  are amplified.  This
amplification is caused by the AO servo system and depends on the type
of temporal  controller used; the SP controller  gives more high-fequency
errors than INT.  The spectra of residual coefficients match very well
their expected shape -- a product  of the input disturbance and the AO
loop error transfer function (Fig.~\ref{fig:pow}).

Quasi-periodic  signals  caused  by  vibrations  are  seen  as  narrow
``lines''  superposed on  the otherwise  smooth spectra  of  input or
residual  distortions.  Such  lines  are  detected in  modes  up to  6
(astigmatism),  but are  not present  in the  higher-order  modes. The
focus coefficient  $a_4$ almost  always has a  line at  47\,Hz. This
frequency is  too high to be caused  by a real focus  variation in the
telescope. It  is suspiciousy close to  the 1/5 of  the loop frequency
and is likely of instrumental origin (we do not see such signal in the
laboratory, however).   On the other  hand,   the 50-Hz
component in the tip and tilt coefficients is real.  This vibration is
caused by the  pick-up noise in the position sensors  of the SOAR fast
tip-tilt mirror,  M3.  This  signal is detected  by the  standard SOAR
guider,  as  well  as  directly  in  the voltages  of  the  M3  drives
(M.~Warner, private  communication). By extracting  the 50-Hz component
from the  SAM loop signal, we   show that motions on  both axes are
co-phased  (Fig.~\ref{fig:tilt}).  The  amplitude of  these vibrations
and relative phase shift between the axes are not constant. On October
2, 2009,  the worst-case  rms amplitude was  19\,mas, the  smallest one --
3\,mas.

As the  typical exposure time  in speckle interferometry is  20\,ms --
one full  period of  the 50-Hz vibration  -- the speckle  structure is
sometimes blurred  and the resolution  does not reach  the diffraction
limit (30\,mas in the visible). If the vibration ellipse is narrow, as
in Fig.~\ref{fig:tilt},  the resulting blur  can mimic a  close binary
with  separation of  20-50\,mas. Such  effect was  indeed seen  in the
previous  speckle  data  obtained  at  SOAR  (TMH10).  It  biases  the
measurements   of   very   close   binaries  in   un-controlled   way.
Modification of  the M3  drive electronics is  in progress.   Once the
50-Hz component is eliminated, there  will be no other sources of fast
blur, making the SOAR  telescope most suitable for diffraction-limited
observations.

\section{Imaging}
\label{sec:img}

Series of short-exposure images were processed in three different ways
(Fig.~\ref{fig:BU172}).  In  all   cases  we  subtracted  the  average
background to remove  the fixed offset and the  dark current which was
non-negligible  in a  small number  of ``hot''  pixes. All  images are
Nyquist-sampled.

{\bf Speckle interferometry} in its standard form is implemented as
described  in  TMH10. We compute the  power spectrum from  all images in
each  series,  remove  the  photon-noise  bias,  filter  out  the  low
frequencies  and  derive by  inverse  Fourier  transform the  average
auto-correlation function  (ACF).  The phase information  is lost, but
the  diffraction-limited resolution  can  be reached  even without  AO
correction.

We found  that  fast wave-front  distortions amplified by
the  SP controller in  closed loop  (cf. Fig.~\ref{fig:pow})  blur the
high-frequency  structure   in  the  20-ms  images   and  degrade  the
resolution in the speckle mode. At the same time this controller gives
similar or smaller overall  wave-front residuals as the INT controller
with a  gain of  0.25; the resolution  in the long-exposure  images is
also a little better with the SP controller.

All  data  on  binary   stars  were  processed  uniformly  by  speckle
interferometry  to derive relative  component positions  and magnitude
differences  (see Sect.~\ref{sec:res}).   For the  analysis of   AO
imaging   performance,  we  de-convolve   the  binary   signature  (as
determined  by the  speckle processing)  and derive  the  point spread
function (PSF).

{\bf  Long-exposure images} were  obtained as  simple averages  of the
data cubes. Typically, the total exposure was about 12\,s (30\,Hz frame
rate, 400  frames). As an option, we  could re-center individual
images before averaging; the images  were shifted by integer number of
pixels as determined by their centroids. Re-centering brings almost no
resolution  gain when  the AO  loop is  closed. Therefore  we restrict
further analysis to long-exposure images. 

{\bf SAA  imaging} is implemented  by selecting a certain  fraction of
the  sharpest  images  in  a  given series  and  co-adding  them  with
re-centering  on  the  brightest   pixel.   In  this  case,  the  most
prominent,  diffraction-limited  speckle is  enhanced  in the  average
image, while  fainter speckles  are averaged out  and contribute  to the
diffuse halo.  As a result,  a PSF with a diffraction-limited core and
a  halo  is  obtained,  reminiscent  of  partially-corrected  AO
images. Combination of partial  AO correction with re-centering on the
brightest pixel gives better results than each of these methods alone.

For the image selection,  we experimented with different criteria such
as second moment or sharpness  and found that selection on the maximum
image intensity  works best, in agreement with  \citet{Smith09}. As to
the fraction of  retained images, it is not  critical.  The resolution
gain is achieved mostly by  re-centering, not by selection.  We co-add
the best 20\% frames to  produce the SAA images.  For faint stars,
the choice of the brightest pixel is affected by the noise, leading to
a sharp  1-pixel spike at the  center; this spike was  replaced by the
average  of  its neighbours for  PSF  evaluation.  In the case
of a binary  star with components of comparable  brightness, the noise
and speckle fluctuations cause the brightest pixel to belong sometimes
to the secondary component.  As a result, a weaker false spike located
symmetrically with respect to the primary component appears in the SAA
image.

\begin{figure}[ht]
\epsscale{1.0}
\plotone{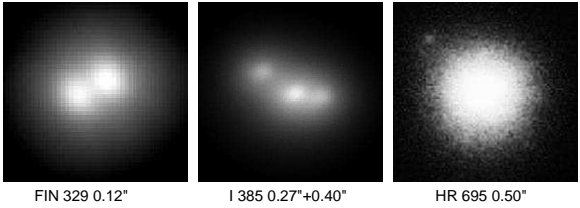} 
\caption{Images of  some multiple  systems in the  $I$ band.   Left: a
  bright  close  pair FIN~329,  long-exposure.   Center: tight  visual
  triple  system I~385 with  comparable separation  between components,
  long  exposure.   Right: astrometric  binary  $\kappa$~For with  a
  faint ($\Delta  m = 3.4$) companion  first resolved at  SOAR in 2007
  (SAA image).
\label{fig:bin}
}
\end{figure}

Figure~\ref{fig:bin}  gives  some  representative images  of  multiple
systems  observed  on October  2,  2009 in  closed  loop  with the  SP
controller.   As expected,  the best  resolution in  the long-exposure
compensated   images  is   reached   in  the   $I$  band   (wavelength
0.8\,$\mu$m), so  most of the data  were obtained in  this filter. The
resolution  gain   over  un-compensated,  seeing-limited   imaging  is
impressive,  even at  shorter wavelengths.  For example,  three images
taken one after  another  in the  $I$, $V$, and $B$
filters  show  the  long-exposure  FWHMs  of  $0.16''$,  $0.30''$  and
$0.44''$ respectively under a $0.80''$ seeing. 

\begin{figure}[ht]
\epsscale{1.0}
\plotone{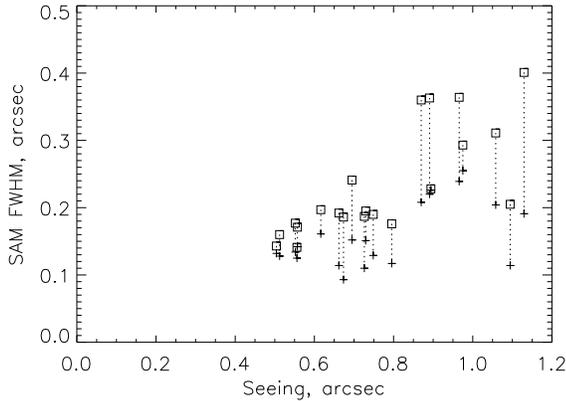} 
\caption{Relation between  atmospheric seeing (horizontal axis) and
  the  FWHM of the  compensated images  in the  case of  long exposure
  (squares) and SAA (crosses connected to squares) processing. Data of
  October 2, 2009, $I$-band.
\label{fig:stat}  }
\end{figure}

The  SAA (co-addition  of  re-centered images)  produces  a FWHM  with
diffraction-limited core even without  AO correction. We obtained such
images from  the data  of 2008-2009  speckle runs at  SOAR recorded  under a
0\farcs5 seeing.  Partial AO  correction helps to concentrale the light
in few bright speckles and thus enhances the range of conditions where
this method works well  \citep{Law09}.  We found  that under
partial  AO  correction  the   SAA  post-processing  yields  a  better
resolution than simple averaging.  However, as the seeing improves and
the AO  correction becomes good,  the additional gain brought  by this
technique  becomes less  evident (Fig.~\ref{fig:stat}).   Average FWHM
for the 22 points in  Fig.~\ref{fig:stat} is 0\farcs25 and 0\farcs18 for
the long-exposure and SAA, respectively.

\begin{figure}[ht]
\epsscale{1.0}
\plotone{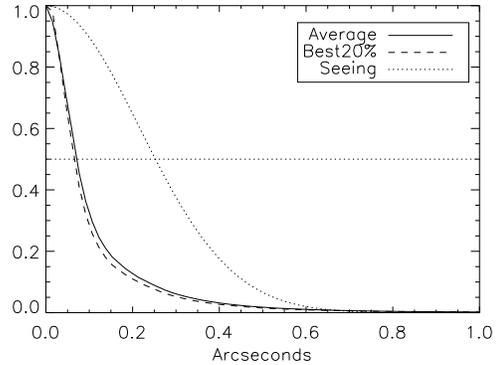} 
\plotone{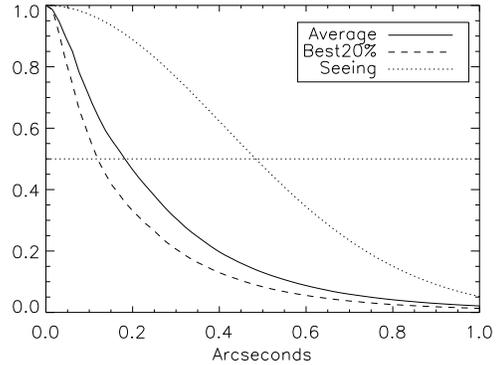} 
\caption{Radial profiles of the PSFs of long-exposure (full lines) and
  SAA  (dashed lines)  images  under good  (top,  $r_0= 0.18$\,m)  and
  mediocre (bottom, $r_0= 0.095$\,m) seeing in the $I$ band.  Gaussian
  PSFs of seeing-limited width are plotted for reference.
\label{fig:PSF}  }
\end{figure}

Radial profiles of  the PSFs under good and  partial AO correction are
plotted in Fig.~\ref{fig:PSF}.  For comparison, Gaussian profiles with
the seeing-limited FWHM $0.98  \lambda/r_0$ are given, using  $r_0$
values estimated form the AO  loop. The shape of the AO-corrected PSFs
is  different  from a Gaussian and resembles  a negative exponent.
Such PSF  is typical for  an AO system like SAM, according to our
simulations.

A detailed comparison of various processing techniques is presented in
Fig.~\ref{fig:BU172}. The  binary star  BU~172 was observed  at zenith
distance $29.5^\circ$ under average seeing ($r_0 = 0.12$\,m to 0.14\,m
from the loop data).  The  first row demonstrates good compensation in
the $I$ band (long-exposure FHWM 0\farcs18) achieved with the INT loop
controller.  The SAA and speckle processing increase the resolution to
the point  where the atmospheric dispersion (47\,mas  in the direction
parallel to the binary) becomes visible.  The dispersion is negligibly
small  (5\,mas)  in  the  narrow-band  H$\alpha$ filter,  but  the  AO
compensation  at   this  longer  wavelength  is  not   so  good  (FWHM
0\farcs33). The  photon flux is  low, so the  SAA method does  not work
very well, while the speckle  processing gives a stronger signal.  The
images  are  elongated perpendicularly  to  the  binary  by the  50-Hz
vibrations (see the insert) of peak-to-peak amplitude 45\,mas, in this
case.  Finally, when the SP controller is used, the images become more
fuzzy because  of the rapid  residual wavefront jitter.   However, the
average image is  even a little sharper (FWHM  0\farcs31) than with the
INT controller.

Although the binary BU~172 is bright, the flux is decreased by $2.1^m$
by the  beamsplitter in  SAM and, further,  by $3.4^m$ by  the narrow-band
H$\alpha$ filter (in comparison with $V$ or $R$ bands). Therefore, the
photon  flux is  equivalent to  a binary  star with  $V =  11.9, 12.1$
observed  in  a wide  band  without  SAM  beamsplitter.  Further  gain  in
sensitivity  by  at least  $\sim  1^m$  is  achievable with  a  better
detector; the  quantum efficiency is  presently only about  0.5, while
the clock injection  charge (about 40 photons per  frame) is a serious
limitation for faint stars.   Considering that this observation is not
yet close to  the limit, there are reasons to  believe that stars down
to $V \sim 15^m$ can  be observed at diffraction-limited resolution in
speckle or lucky mode with SAM.

\section{Results on multiple stars}
\label{sec:res}


\subsection{Binary-star measurements}

Relative positions and magnitude difference in close binary stars were
measured by speckle interferometry, following the same procedure as in
TMH10.  We took  advantage of that work to  calibrate new results from
SAM on common wide systems and found that the pixel scale remained the
same  within  errors, 15.23\,mas.   An  offset  in  position angle  of
$6.0^\circ$  has been determined  and corrected.   For the  most part,
we measured  bright and  well-known systems,  although 
some recently discovered close pairs were also observed.

Table~\ref{tab:double} lists  122 measurements of  75  resolved
binary   stars.   Its   columns  contain   (1)  the   WDS  \citep{WDS}
designation, (2) the ``discoverer  designation" as adopted in the WDS,
(3)  an  alternative name,  mostly  from  the  {\it Hipparcos} catalog,  (4)
Besselian epoch  of observation, (5) filter, (6)  number of individual
data  cubes, (7,8)  position angle  $\theta$ in  degrees  and internal
measurement  error in tangential  direction $\rho  \sigma_{\theta}$ in
mas, (9,10)  separation $\rho$ and its  internal error $\sigma_{\rho}$
in mas, and (11) magnitude  difference $\Delta m$. An asterisk follows
if $\Delta  m$ is determined  from the resolved photometry  (see TMH10
for details). A colon indicates that the data are noisy and $\Delta m$
is likely over-estimated.   Note that in the cases  of multiple stars,
the positions and photometry  refer to the pairings between individual
stars, not with photo-centers of sub-systems.

For   stars  with   known  orbital   elements,  columns   (12--14)  of
Table~\ref{tab:double}  list the residuals  to the  ephemeris position
and the reference  to the orbit from the  {\it 6$^{th}$ Orbit Catalog}
\citep{VB6}. Some  measurements from this work were  used to calculate
improved orbits by \citet{Mason10}; these orbits are referenced as Msn2010 in 
Table~\ref{tab:double}. 

In most cases we see binary  companions in the SAA images and can thus
resolve the $180^\circ$ ambiguity inherent to the speckle method.  For
a  few pairs  where this  was not  possible, the  position  angles are
preceded by the letter A (ambiguous).

\subsection{Comments on individual systems}

\begin{figure}[ht]
\epsscale{1.0}
\plotone{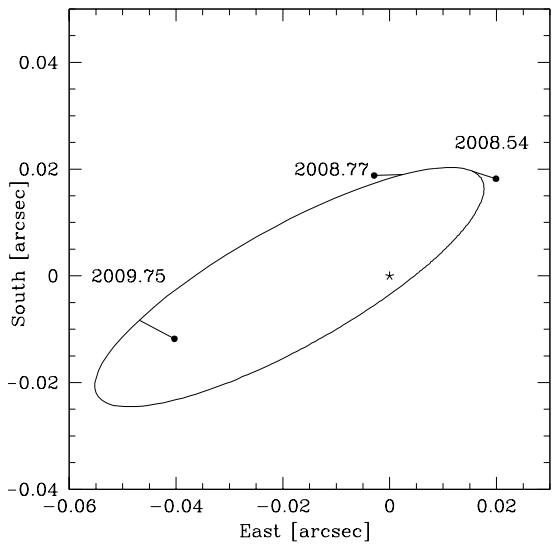} 
\caption{Preliminary visual orbit of HIP~9774 Aa,Ab = WSI~71.
\label{fig:orb}  }
\end{figure}

00522$-$2237 =  STN 3 has a new faint tertiary component  D at 0\farcs57
from B with $\Delta I = 4^m$. As the separation in the wide pair AB is
only $2''$,  the triple  system appears to  be close to  the dynamical
stability limit. We  cannot exclude that the new  companion is just an
unrelated  background source projecting  on the  AB.  The  companion C
listed in the WDS is in fact  such a background source, as evidenced by
its  relative motion  since 1897.  However,  the actual separation  of AC  is
$>10''$, so the C companion was well outside the field.

02057$-$2423 =  HIP~9744 is  a close visual  triple star consisting  of the
0\farcs8 pair AB  = I~454  and  the inner  spectroscopic binary  Aa,Ab
\citep{TS02}. It  forms a quadruple or quintuple  system together with
HIP~9769 which has common radial velocity, parallax, and proper motion
with HIP~9774, at $56''$ separation.

We determined a preliminary  spectroscopic orbit of the Aa,Ab sub-system
with a period of 2.6~y and eccentricity 0.76. The estimated semi-major
axis of this pair is 43\,mas,  indicating that it can be resolvable at
4-m telescopes.  First resolution of  Aa,Ab (named on this occasion
WSI~71) was achieved  in July 2008, shortly after  passage through the
periastron (TMH10).  The pair  was resolved again  in October  2008 at
SOAR and observed  also on October 2, 2009 with SAM.  By that time the
components opened up to 42\,mas.

An attempt to  combine the three position measurements  with the known
spectroscopic  orbit  has shown  that  the  separations and  magnitude
differences   in  the  published   speckle  observations   are  likely
over-estimated. They were determined  jointly with the position of the
wide pair  Aa,B by fitting  a triple-star model.  We  re-processed all
speckle data by imposing a constraint $\Delta m = 0.30$, as determined
from  the relative  intensity  of component's  lines  in the  resolved
spectra. The presence  of the wide companion B  was ignored by fitting
only the close binary.   This reprocessing gives closer separations of
Aa,Ab,  as  expected (Table~\ref{tab:orbit}).   No  formal errors  are
given  because measurements of  such close  binaries can  be seriously
biased by vibrations (see above).

\setcounter{table}{1}
\begin{table}
\center
\tabletypesize{\scriptsize}                                                                                                         
\caption{Measurements of HIP~9774 Aa,Ab
\label{tab:orbit} }
\medskip                                                                             

\begin{tabular}{ ccc cc }        
\hline
\hline
 Epoch  & $\theta$ &    $\rho$  & [O$-$C]$_{\theta}$ & [O$-$C]$_{\rho}$ \\  
        & (deg)    &  ($''$)    & (deg) & ($''$) \\                   
\hline
 2008.5406 &  132.4 &  0.027 & $-$9.9 &  0.002  \\
 2008.7700 &  188.7 &  0.019 &   15.8 &  0.000   \\
 2009.7558 &  286.3 &  0.042 &    5.1 &  $-$0.007 \\
\hline
\end{tabular}         
\end{table}  

Figure~\ref{fig:orb}   shows  a   preliminary   interferometric  orbit
obtained   by   fitting  jointly   radial   velocities  and   resolved
measurements.   Clearly,  more  data   are  needed  for  a  definitive
solution.   However,  plausible  values  of the  visual  elements  are
obtained:  semi-major axis  $a  = 58$\,mas,  position  angle of  nodes
$\Omega = 125^\circ$ and inclination $i = 77^\circ$.

Speckle  interferometry is the  {\it only}  method for  resolving such
close binaries.  Adaptive optics at 8-m telescopes has somewhat poorer
resolution (diffraction limit $\lambda/D  = 32$\,mas in the $J$ band),
whereas  long-baseline   interferometers  lack  the   sensitivity  and
productivity to follow such binaries.


22116$-$3428  =  HIP 109561  is  a  visual  triple consisting  of  the
0\farcs4 pair  AB =  BU~769 and  the inner pair  Aa,Ab =  CHR~230.  We
marginally resolved  the inner sub-system at almost  the same position
angle as measured in TMH10 (its estimated period is 100\,y). Note that
component A  is a K1III giant  and B is  a blue star, possibly  of A5V
spectral type. Therefore  the magnitude difference of AB  is larger at
longer wavelengths.

\begin{figure}[ht]
\epsscale{1.0}
\plotone{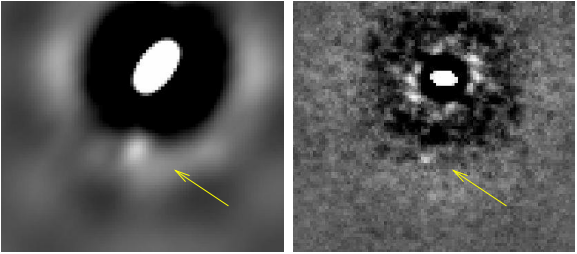} 
\caption{Fragments of the ACFs of $\zeta$  Aqr B in the $I$ (left) and
  H$\alpha$ (right)  bands.  The arrows mark  the tentatively resolved
  astrometric companion Bb at 0\farcs4.
\label{fig:ZetAqr}  }
\end{figure}


\begin{figure}[ht]
\epsscale{1.0}
\plotone{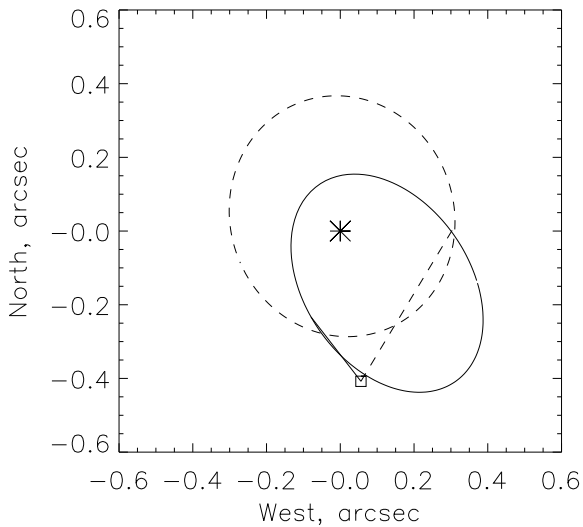} 
\caption{Observed position of $\zeta$ Aqr Ba,Bb (square) in comparison
  with  astrometric   orbits  of  \citet[][full line]{Heintz84}   and
  \citet[][dashed line]{Scardia09}. For  both astrometric  orbits, the
  position angle is  inverted and the semi-major axis  $a = $0\farcs34 is
  adopted,  as appropriate  for  direct resolution.   The companion  is
  connected to its predicted positions on both orbits.
\label{fig:orbits}  }
\end{figure}

22288$-$0001 =  $\zeta$ Aqr  B = HR  8558 is  a secondary in  the wide
visual pair  AB = STF 2909  with components of  similar spectral types
and magnitudes (F3V and F6IV, $V = 4.42; \; 4.51$) observed since 1821.
Both stars are  above the Main Sequence and  have fast axial rotation.
A ``wavy'' motion  of AB suggested that there  is an inner sub-system.
\citet{Strand42}  published  the   first  astrometric  orbit  of  this
sub-system,  attributing  it  to  the component  B.   Several  authors
confirmed subsequently the 25-y astrometric wave and refined both outer
and inner orbits; the  latest are  papers by \citet{Heintz84}
and \citet{Scardia09}.  The first direct resolution of the astrometric
companion in the $K$ and  $L$ bands was reported by \citet{McCarthy82}
who  fitted one-dimensional  scans  in the  North-South  direction to  a
triple-star  model. No  further  resolved observations  of Ba,Bb  were
published to date.   The component  A was  also resolved  with speckle  into a
0\farcs064  binary by  \citet{EW79},  but never  confirmed ever  since,
despite    several    speckle    measurements    of    AB    at    4-m
telescopes. Interestingly, \citet{Heintz84} dismisses both resolutions
as  bogus  and attributes  the  25-y sub-system  to  A,  based on  the
astrometry of A and B.

Despite  the controversy,  the existence  of a  25-y  sub-system makes
little  doubt.   Let us assume  the mass  of  the  Ba  star $M_{\rm  Ba}  =
1.5\,M_\odot$ in  agreement with its  spectral type and  the dynamical
mass  sum derived  from the  AB  orbit and  adopt the mass   sum  of
1.9\,$M_\odot$ for Bab. Then its period  25.8\,y and the third Kepler's law
lead  to a  semi-major axis  of  10.8\,AU, or  $a =  $0\farcs34 with  the
Hipparcos parallax of 31.5\,mas.   The comparison with the astrometric
semi-major axis $\alpha =  75.5$\,mas \citep{Heintz84} leads to a mass
ratio  $q  =  M_{\rm  Bb}/  M_{\rm  Ba} =  0.28$  and  $M_{\rm  Bb}  =
0.4\,M_\odot$,  in  broad agreement  with  other published  estimates.   The
unseen component  could be an M0V  dwarf.  In this  case the magnitude
difference  with  Ba in  the  $I$ band  would  be  around $5^m$.   The
estimated radial-velocity variation of  Ba is $K_1 \sim 1.3$\,km/s; it
has never been detected, possibly because of the fast axial rotation.

We see a faint ($\Delta m \sim 5^m$) companion to B in four data cubes
-- two in  the band $I$ and two  in H$\alpha$ (Fig.~\ref{fig:ZetAqr}).
This resolution  should be considered as tentative  until confirmed by
further measurements.   It would be  relatively easy to detect  such a
companion with AO in the  infrared.  The component A has also been observed
with SAM  and found unresolved.

The observed position  of Ba,Bb, if it is real, does  not match well the
two  recent astrometric  orbits,  which also  differ  from each  other
substantially  (Fig.~\ref{fig:orbits}). In both  cases we  adopted the
semi-major axis  of 0\farcs34.  If  we correct the periastron  epoch in
the orbit of  \citet{Heintz84} from the original 2007.0  to 2004.0, it
would agree with our observation.



\section{Conclusions and outlook}
\label{sec:concl}

To summarize,  the first tests of SAM  in its NGS mode  have shown the
following:
\begin{itemize}
\item
The AO system works as expected when closed on natural guide
stars. The FWHM resolution of AO-corrected long-exposure images can be
better than 0\farcs2 in the $I$-band and is significantly improved with
respect to the seeing across the visible spectrum. 

\item
Fine speckle structure  in the images with 20-ms  exposure time can be
blurred by the 50-Hz vibration originating in the SOAR tip-tilt mirror
(soon to  be improved)  and by the  amplification of  fast atmospheric
distortions  in  the  AO  loop  using  the  SP  controller.  A  softer
controller is preferable for diffraction-limited imaging.

\item
The  SAA  (``lucky'')  processing  of  the  data  cubes  improves  the
resolution for  bright sources  under partial compensation.   It gives
little additional gain  under good AO correction and  is less powerful
than the speckle processing in the photon-starved regime.

\item
Both speckle and SAA work well for  stars as faint as $V \sim 12^m$ if
all available photons  are used by the current  HRCam.  A further gain
of 1-2 magnitudes  could be achievable by using state-of-the-art EM
CCD with a higher quantum efficiency and lower charge injection noise.

\end{itemize}

These  first  results  suffer  from small  statistics  and  incomplete
sampling  of atmospheric  conditions; more  experiments are  needed to
substantiate and extend these findings. 


Methods  like  speckle and  SAA  imaging,  enhanced  by   partial  AO
correction, open a unique high-resolution niche. A non-exhaustive list
of potential science programs includes:

\begin{itemize}
\item
Followup of  close nearby binaries and determination  of their orbits,
leading  to  mass  measurement  (including  pre-main-sequence  stars),
statistics of orbital elements, dynamics of multiple systems, etc.

\item
Surveys  of  various  stellar  polulations  for  binarity  delivering
observational constraints on star formation.

\item
Optical imaging of solar-system objects (planetary satellites, asteroids), see
e.g. \citep{Mercury}. 

\item
Studies of dense stellar aggregates (globular clusters, young groups
like R~136a). 

\item
Resolved emission details around young stars (jets, disks)

\end{itemize}

Further interesting  results are expected from the  combination of the
HRCam  with   SAM.   In  a  more  distant   perspective,  a  dedicated
high-resolution  instrument  will be  worth  considering.   As the  AO
technology  becomes  mature and  accessible,  we  envision building  a
small-field camera with  integrated compact AO system.  The light from
the  object (or  its  bright  stellar component)  will  be divided  in
wavelength between the  science channel and the WFS,  e.g.  by sensing
the beam reflected from the science filter.  An advanced WFS with a EM
CCD would  be able to operate  on much fainter stars  than is possible
with SAM now.  In fact, even  the faintest stars will be useful.  When
the  number   of  photons  is  not  sufficient   for  fast  turbulence
correction,  the AO system  will work  in the  slow mode  to compensate
residual telescope aberrations  (mostly defocus and tilts), delivering
to the speckle camera truly seeing-limited images.

\acknowledgments

 We  thank the  SOAR telescope  team for their essential help with the 
installation of the SAM instrument and observations.
This research has  made use of the CDS and SIMBAD  services and of the
Washington  Double   Star  Catalog   maintained  at  the   U.S.  Naval
Observatory.


\clearpage

\setcounter{table}{0}
\begin{deluxetable}{l l l  ccc  rc cc l r r l }                                                                                                                                
\tabletypesize{\scriptsize}                                                                                                                                                     
\rotate                                                                                                                                                                         
\tablecaption{Measurements of binary stars at SOAR                                                                                                                                
\label{tab:double} }                                                                                                                                                            
\tablewidth{0pt}                                                                                                                                                                
\tablehead{                                                                                                                                                                     
\colhead{WDS} & \colhead{Discoverer} & \colhead{Other} & \colhead{Epoch} & \colhead{Filt} & \colhead{N} & \colhead{$\theta$} & \colhead{$\rho \sigma_{\theta}$} &               
\colhead{$\rho$} & \colhead{$\sigma \rho$} & \colhead{$\Delta m$} & \colhead{[O$-$C]$_{\theta}$} & \colhead{[O$-$C]$_{\rho}$} & \colhead{Reference} \\         
\colhead{(2000)} & \colhead{Designation} & \colhead{name} & +2000 & & & \colhead{(deg)} & (mas) & ($''$) & (mas) & (mag) & \colhead{(deg)} & \colhead{($''$)}                   
& \colhead{code$^*$}  }   
\startdata     
00143$-$2732 &	HDS 33 &	HIP 1144 &	9.6709 &	I &	2 &	310.8 &	0.3 &	0.1674 &	0.9 &	0.9    &  &   &   \\ 
00284$-$2020 &	B 1909 &	HIP 2237 &	9.6709 &	I &	3 &	307.8 &	0.5 &	0.1888 &	0.5 &	0.2   &	$-$3.8 &	0.003 &	Sod1999  \\ 
00315$-$6257 &	I 260 CD &	HIP 2487 &	9.6711 &	y &	2 &	242.4 &	0.4 &	0.2529 &	0.3 &	1.5   &	5.6 &	$-$0.017 &	Msn2001c  \\ 
00345$-$0433 &	D 2 AB &	HIP 2713 &	9.6708 &	I &	2 &	A207.9&	0.4 &	0.0610 &	0.1 &	0.8   & 109.9 &	$-$0.223 &	Sey2002  \\ 
            &	       &	           &	9.6708 &	R &	2 &	A207.8&	1.0 &	0.0604 &	0.9 &	0.1   &	109.8 &	$-$0.223 &	Sey2002  \\ 
            &	       &	           &	9.7556 &	I &	6 &	A204.7&	5.3 &	0.0574 &	1.2 &	0.9   &	106.6 &	$-$0.228 &	Sey2002  \\ 
00352$-$0336 &	HO 212 AB &	HIP 2762 &	9.6708 &	I &	2 &	229.6 &	0.1 &	0.2345 &	0.4 &	1.2   &	$-$0.7 &	0.000 &	Msn2005  \\ 
            &	       &	           &	9.6708 &	y &	2 &	229.5 &	0.4 &	0.2344 &	1.0 &	1.4   &	$-$0.7 &	0.000 &	Msn2005  \\ 
            &	       &	           &	9.7556 &	I &	4 &	231.7 &	0.5 &	0.2387 &	0.2 &	1.1   &	$-$0.4 &	0.000 &	Msn2005  \\ 
00373$-$2446 &	BU 395 &	HIP 2941 &	9.6709 &	I &	4 &	95.0 &	0.1 &	0.4627 &	0.1 &	0.4   &	2.8 &	0.017 &	Pbx2000b  \\ 
            &	       &	           &	9.7557 &	I &	2 &	95.6 &	0.1 &	0.4727 &	0.3 &	0.2 * &	2.9 &	0.018 &	Pbx2000b  \\ 
00427$-$6537 &	I 440 &	HIP 3351 &	9.6711 &	I &	2 &	269.4 &	0.2 &	0.4023 &	0.3 &	0.7   &	1.1 &	$-$0.084 &	Lin2004a  \\ 
00522$-$2237 &	STN 3 AB &	HIP 4072 &	9.6709 &	I &	3 &	243.1 &	1.4 &	1.9853 &	2.3 &	0.7 *  &  &   &   \\ 
00522$-$2237 &	STN 3 BD &	HIP 4072 &	9.6709 &	I &	3 &	336.9 &	5.9 &	0.5743 &	3.3 &	4.0    &  &   &   \\ 
01061$-$4643 &	SLR 1 AB &	HIP 5165 &	9.6711 &	y &	2 &	113.0 &	0.1 &	0.4050 &	0.1 &	0.5   &	$-$29.2 &	$-$0.042 &	Ary2001a  \\ 
01078$-$4129 &	RST 3352 &	HIP 5300 &	9.6711 &	I &	2 &	156.1 &	0.2 &	0.1412 &	0.3 &	1.0   &	1.9 &	$-$0.003 &	Sod1999  \\ 
            &	       &	           &	9.6711 &	y &	2 &	156.1 &	0.4 &	0.1408 &	0.3 &	1.3   &	2.0 &	$-$0.004 &	Sod1999  \\ 
01084$-$5515 &	RST 1205 AB &	HIP 5348 &	9.6711 &	y &	2 &	110.0 &	1.5 &	0.5486 &	0.4 &	2.7   &	$-$2.3 &	0.033 &	Lin2004a  \\ 
01094$-$5636 &	HU 1342 &	HIP 5428 &	9.6711 &	I &	3 &	328.8 &	0.2 &	0.3746 &	0.7 &	0.6   &	$-$1.8 &	$-$0.018 &	Hei1984a  \\ 
01158$-$6853 &	I 27 CD &	HIP 5842 &	9.6711 &	I &	2 &	312.5 &	0.3 &	1.0540 &	0.2 &	0.5 * &	2.0 &	$-$0.034 &	Sod1999  \\ 
01198$-$0031 &	FIN 337 BC &	HIP 6226 &	9.6710 &	I &	2 &	32.2  &	1.4 &	0.1337 &	1.7 &	0.8   &	$-$3.8 &	0.003 &	Msn2010   \\ 
01198$-$0031 &	STF 113 AB &	HIP 6226 &	9.6710 &	I &	2 &	19.1 &	3.4 &	1.6783 &	0.3 &	2.4   &	 &	 &  \\ 
01220$-$6943 &	I 263  &	HIP 6377 &	9.6711 &	I &	2 &	279.1 &	0.2 &	0.4271 &	0.2 &	0.8   &	5.3 &	$-$0.099 &	Msn1999a  \\ 
01243$-$0655 &	BU 1163 &	HIP 6564 &	9.6710 &	I &	3 &	218.0 &	0.2 &	0.3227 &	0.1 &	0.4   &	$-$0.4 &	$-$0.002 &	Sod1999  \\ 
01334$-$4354 &	HDS 205 &	HIP 7254 &	9.6711 &	I &	2 &	195.8 &	0.2 &	0.1322 &	0.7 &	0.9    &  &   &   \\ 
01350$-$2955 &	BU 1000 AC &	HIP 7372 &	9.6709 &	I &	2 &	351.1 &	2.1 &	1.8128 &	0.9 &	3.2   &	$-$17.8 &	0.237 &	Nwb1969a  \\ 
01350$-$2955 &	DAW 31 AB &	HIP 7372 &	9.6709 &	I &	2 &	40.7 &	2.6 &	0.1366 &	3.3 &	0.5   &	$-$7.0 &	$-$0.005 &	Msn1999c  \\ 
01361$-$2954 &	HJ 3447 &	HIP 7463 &	9.6709 &	I &	2 &	182.5 &	0.3 &	0.7729 &	0.1 &	1.4   &	$-$2.9 &	$-$0.038 &	Cve2006e  \\ 
01376$-$0924 &	KUI 7 &	HIP 7580 &	9.6710 &	y &	1 &	112.5 &	0.2 &	0.0771 &	0.2 &	1.2 : &	3.7 &	0.002 &	Tok1993  \\ 
            &	       &	           &	9.6710 &	I &	2 &	111.9 &	0.4 &	0.0786 &	0.7 &	1.0   &	3.1 &	0.004 &	Tok1993  \\ 
01417$-$1119 &	STF 147 &	HIP 7916 &	9.6709 &	I &	2 &	120.2 &	0.1 &	0.1072 &	0.2 &	1.0    &  &   &   \\ 
            &	       &	           &	9.6710 &	y &	2 &	120.4 &	0.1 &	0.1065 &	0.5 &	1.1    &  &   &   \\ 
02022$-$2402 &	HDS 272 AB &	HIP 9497 &	9.6710 &	I &	3 &	340.6 &	0.8 &	0.5861 &	1.6 &	3.4    &  &   &   \\ 
            &	       &	           &	9.6710 &	R &	2 &	340.6 &	0.8 &	0.5801 &	0.3 &	4.4    &  &   &   \\ 
            &	       &	           &	9.7557 &	I &	3 &	340.2 &	1.1 &	0.5851 &	2.3 &	3.5    &  &   &   \\ 
02022$-$2402 &	TOK 41 Ba,Bb &	HIP 9497 &	9.6709 &	I &	3 &	162.2 &	0.3 &	0.1018 &	1.0 &	0.1    &  &   &   \\ 
            &	       &	           &	9.6710 &	R &	2 &	162.7 &	1.4 &	0.0948 &	2.9 &	0.0    &  &   &   \\ 
            &	       &	           &	9.7557 &	I &	3 &	160.1 &	4.0 &	0.1007 &	2.5 &	0.0    &  &   &   \\ 
02057$-$2422 &	WSI 71 Aa,Ab &	HIP 9774 &	9.7557 &	I &	2 &	276.1 &	0.8 &	0.0462 &	8.2 &	1.4 :  &  &   &   \\ 
            &	       &	           &	9.7558 &	I &	2 &	283.7 &	2.3 &	0.0553 &	0.1 &	1.4 :  &  &   &   \\ 
            &	       &	           &	9.7559 &	I &	2 &	287.2 &	8.3 &	0.0525 &	1.4 &	1.5 :  &  &   &   \\ 
02057$-$2423 &	I 454 AB &	HIP 9774 &	9.7557 &	I &	2 &	154.4 &	0.1 &	0.8204 &	1.1 &	1.9 :  &  &   &   \\ 
            &	       &	           &	9.7558 &	I &	2 &	154.6 &	0.3 &	0.8173 &	0.1 &	2.2 :  &  &   &   \\ 
            &	       &	           &	9.7559 &	I &	2 &	154.7 &	0.6 &	0.8185 &	3.1 &	2.0 :  &  &   &   \\ 
02225$-$2349 &	TOK 40 &	HIP 11072 &	9.6709 &	I &	2 &	125.3 &	1.2 &	0.4969 &	1.2 &	3.8    &  &   &   \\ 
            &	       &	           &	9.6709 &	R &	2 &	124.8 &	4.6 &	0.4997 &	4.6 &	4.5    &  &   &   \\ 
            &	       &	           &	9.7558 &	I &	5 &	126.1 &	0.9 &	0.4966 &	1.2 &	3.6 *  &  &   &   \\ 
02396$-$1152 &	FIN 312 &	HIP 12390 &	9.6710 &	y &	2 &	131.1 &	0.3 &	0.0762 &	0.2 &	0.8   &	7.6 &	$-$0.003 &	Sod1999  \\ 
02442$-$2530 &	FIN 379 Aa,Ab &	HIP 12780 &	9.6709 &	I &	2 &	302.2 &	0.4 &	0.0648 &	0.7 &	0.2   &	$-$41.6 &	$-$0.279 &	Hng2005  \\ 
            &	       &	           &	9.7557 &	I &	3 &	310.1 &	0.9 &	0.0727 &	2.8 &	0.5   &	$-$33.9 &	$-$0.270 &	Hng2005  \\ 
            &	       &	           &	9.7558 &	V &	1 &	312.2 &	0.5 &	0.0596 &	0.5 &	0.7 : &	$-$31.8 &	$-$0.283 &	Hng2005  \\ 
02460$-$0457 &	BU 83  &	HIP 12912 &	9.6710 &	I &	2 &	14.8 &	0.4 &	0.9560 &	0.4 &	1.6 * &	1.3 &	0.130 &	Ole2002d  \\ 
02572$-$2458 &	BEU 4 Ca,Cb &	HIP 13769 &	9.6709 &	I &	2 &	A352.4&	0.3 &	0.0818 &	0.9 &	0.8   &	10.6 &	$-$0.985 &	Sca2002c  \\ 
17031$-$5314 &	HDS 2412 Aa,Ab&	HIP 83431 &	9.7551 &	I &	4 &	199.4 &	0.7 &	0.5939 &	0.9 &	2.7 *  &  &   &   \\ 
            &	       &	           &	9.7551 &	H$\alpha$ &2 &	199.3 &	0.7 &	0.5925 &	1.3 &	2.8    &  &   &   \\ 
17195$-$5004 &	FIN 356 &	HIP 84759 &	9.7551 &	I &	2 &	A84.7&	0.2 &	0.0698 &	0.4 &	0.0    &  &   &   \\ 
            &	       &	           &	9.7552 &	H$\alpha$ &	2 &	A91.0&	0.2 &	0.0647 &	0.7 &	0.4 :  &  &   &   \\ 
            &	       &	           &	9.7564 &	H$\alpha$ &	2 &	A89.9&	2.2 &	0.0645 &	0.5 &	0.7 :  &  &   &   \\ 
17248$-$5913 &	I 385 AB &	HIP 85216 &	9.7551 &	I &	3 &	121.3 &	2.3 &	0.3984 &	4.8 &	0.8    &  &   &   \\ 
17248$-$5913 &	WSI 87 AD &	HIP 85216 &	9.7551 &	I &	3 &	269.4 &	4.3 &	0.2687 &	7.7 &	0.7    &  &   &   \\ 
18031$-$0811 &	STF 2262 AB &	HIP 88404 &	9.6702 &	I &	2 &	285.2 &	0.8 &	1.6053 &	0.5 &	0.7 * &	0.5 &	$-$0.020 &	Sod1999  \\ 
            &	       &	           &	9.6702 &	R &	6 &	285.2 &	1.2 &	1.6024 &	1.3 &	0.7 * &	0.5 &	$-$0.023 &	Sod1999  \\ 
18112$-$1951 &	BU 132 &	HIP 89114 &	9.6702 &	I &	1 &	8.4 &	1.3 &	1.4047 &	1.3 &	1.3 :  &  &   &   \\ 
18594$-$1250 &	KUI 89 &	HIP 93225 &	9.6703 &	I &	2 &	142.4 &	0.4 &	0.2159 &	0.3 &	0.9   &	12.4 &	0.019 &	Msn1999c  \\ 
            &	       &	           &	9.6703 &	R &	2 &	142.3 &	0.4 &	0.2173 &	0.4 &	1.0   &	12.3 &	0.020 &	Msn1999c  \\ 
            &	       &	           &	9.6703 &	I &	8 &	142.4 &	0.8 &	0.2163 &	0.8 &	0.9   &	12.4 &	0.019 &	Msn1999c  \\ 
20311$-$1503 &	FIN 336 &	HIP 101221 &	9.6705 &	I &	4 &	304.7 &	0.4 &	0.1528 &	0.6 &	1.5   &	 $-$1.9 &	0.000 &	Msn2010  \\ 
            &	       &	           &	9.6705 &	R &	2 &	304.8 &	0.8 &	0.1538 &	0.6 &	1.3   &	 $-$1.8 &	0.001 &	Msn2010  \\ 
20393$-$1457 &	HU 200 AB &	HIP 101923 &	9.6706 &	R &	2 &	118.1 &	0.3 &	0.3313 &	0.3 &	0.6   &	$-$2.1 &	0.006 &	Hei1998  \\ 
            &	       &	           &	9.6706 &	I &	2 &	118.1 &	0.2 &	0.3318 &	0.2 &	0.5   &	$-$2.1 &	0.006 &	Hei1998  \\ 
20514$-$0538 &	STF 2729 AB &	HIP 102945 &	9.6706 &	I &	2 &	25.1 &	0.4 &	0.8567 &	0.7 &	1.5   &	$-$1.3 &	0.080 &	Hei1998  \\ 
20527$-$0859 &	MCA 64 &	HIP 103045 &	9.6706 &	y &	2 &	A137.3&	0.3 &	0.0509 &	0.8 &	0.7 :  &  &   &   \\ 
            &	       &	           &	9.6706 &	I &	2 &	A137.2&	0.1 &	0.0525 &	0.4 &	0.4    &  &   &   \\ 
            &	       &	           &	9.6706 &	R &	2 &	A137.8&	0.4 &	0.0498 &	0.6 &	0.1    &  &   &   \\ 
21041$-$0549 &	MCA 66 Aa,Ab &	HIP 103981 &	9.6706 &	I &	2 &	 62.1 &	0.8 &	0.2705 &	0.7 &	3.4    &  &   &   \\ 
21041$-$0549 &	STF 2745 AB &	HIP 103981 &	9.6706 &	I &	2 &	196.5 &	3.4 &	2.4603 &	10.1 &	5.3    &  &   &   \\ 
21044$-$1951 &	FIN 328 &	HIP 104019 &	9.6707 &	I &	2 &	338.6 &	0.1 &	0.3098 &	0.1 &	1.9   &	 176.5 &	0.009 &	Msn1999a  \\ 
            &	       &	           &	9.6707 &	y &	2 &	338.6 &	0.3 &	0.3095 &	0.5 &	2.3   &	 176.5 &	0.009 &	Msn1999a  \\ 
21074$-$0814 &	BU 368 AB &	HIP 104272 &	9.6706 &	I &	2 &	284.8 &	0.5 &	0.1136 &	0.9 &	0.9   &	0.4 &	0.049 &	Pal2005b  \\ 
            &	       &	           &	9.6707 &	R &	2 &	286.5 &	0.8 &	0.1149 &	2.0 &	0.9   &	2.1 &	0.050 &	Pal2005b  \\ 
21114$-$5220 &	HU 1626 &	HIP 104604 &	9.6678 &	y &	2 &	117.1 &	1.1 &	1.1015 &	1.0 &	1.5 * &	$-$0.1 &	0.005 &	Sey2002  \\ 
            &	       &	           &	9.6679 &	R &	3 &	117.1 &	1.0 &	1.1027 &	1.2 &	1.3 * &	$-$0.0 &	0.006 &	Sey2002  \\ 
21158$-$5316 &	FIN 329 &	HIP 104978 &	9.7552 &	I &	2 &     236.7 &	0.9 &	0.1233 &	0.5 &	0.1   &	$-$118.2 &	0.064 &	Hei1973b  \\ 
            &	       &	           &	9.7552 &	H$\alpha$ &	4 &   236.7 &	0.9 &	0.1242 &	0.5 &	0.4 : &	$-$118.2 &	0.064 &	Hei1973b  \\ 
21274$-$0701 &	HDS 3053 &	HIP 105947 &	9.6706 &	I &	2 &	A159.0&	0.9 &	0.1951 &	2.4 &	1.5 : &	0.6 &	0.0000 &	Msn2010  \\ 
21552$-$6153 &	HDO 296 &	HIP 108195 &	9.7552 &	I &	4 &	105.9 &	0.3 &	0.3322 &	0.2 &	0.2   &	$-$8.1 &	0.042 &	Fin1969c  \\ 
            &	       &	           &	9.7552 &	H$\alpha$ &	1 &	105.8 &	0.4 &	0.3337 &	0.4 &	0.6 : &	$-$8.2 &	0.044 &	Fin1969c  \\ 
21579$-$5500 &	FIN 307 &	HIP 108431 &	9.6678 &	y &	10 &	75.2 &	0.5 &	0.1189 &	1.0 &	0.6   &	0.5 &	0.018 &	Chu1965  \\ 
22116$-$3428 &	BU 769 AB &	HIP 109561 &	9.6679 &	I &	7 &	355.1 &	0.2 &	0.8399 &	0.6 &	1.8 *  &  &   &   \\ 
            &	       &	           &	9.6679 &	R &	1 &	355.0 &	0.4 &	0.8427 &	0.4 &	1.4 *  &  &   &   \\ 
            &	       &	           &	9.6679 &	y &	2 &	355.2 &	0.5 &	0.8427 &	2.2 &	1.4 :  &  &   &   \\ 
22116$-$3428 &	CHR 230 Aa,Ab &	HIP 109561 &	9.6679 &	y &	2 &	128.8 &	2.4 &	0.0284 &	0.9 &	2.3 :  &  &   &   \\ 
22152$-$0535 &	A 2599 AB &	HIP 109874 &	9.7553 &	I &	2 &	279.6 &	2.4 &	0.6999 &	0.8 &	3.2    &  &   &   \\ 
            &	       &	           &	9.7553 &	H$\alpha$ &	2 &	279.9 &	3.1 &	0.6997 &	4.1 &	3.1 :  &  &   &   \\ 
22161$-$0705 &	HDS 3158 &	HIP 109951 &	9.7553 &	I &	2 &	108.6 &	1.4 &	0.3985 &	0.6 &	1.3 :  &  &   &   \\ 
22241$-$0450 &	BU 172 AB &	HIP 110578 &	9.6707 &	I &	3 &	39.5 &	0.3 &	0.4067 &	0.3 &	0.5   &	$-$2.1 &	$-$0.005 &	Doc2007d  \\ 
            &	       &	           &	9.7552 &	I &	5 &	39.4 &	0.1 &	0.4076 &	0.1 &	0.3   &	$-$2.1 &	$-$0.005 &	Doc2007d  \\ 
            &	       &	           &	9.7552 &	H$\alpha$ &	2 &	39.6 &	2.9 &	0.4081 &	1.1 &	0.8 : &	$-$1.9 &	$-$0.005 &	Doc2007d  \\ 
22266$-$1645 &	SHJ 345 AB &	HIP 110778 &	9.7556 &	I &	4 &	38.8 &	0.7 &	1.3102 &	0.6 &	0.1 * &	0.9 &	$-$0.010 &	Hle1994  \\ 
22288$-$0001 &	$\zeta$ Aqr Ba,Bb&	HIP 110960 &	9.7552 &	I  &	2 &    191.8 & 11.3 &	0.4058 &	2.5 &	4.7    &  &   &   \\ 
             &	              &            &	9.7552 &	H$\alpha$ &2 & 187.8 &	5.9 &	0.4116 &	2.4 &	5.2    &  &   &   \\ 
22384$-$0754 &	A 2695 &	HIP 111761 &	9.6707 &	I &	2 &	254.7 &	2.4 &	0.1024 &	0.7 &	2.0   &	$-$29.8 &	$-$0.011 &	Ole2004a  \\ 
            &	       &	           &	9.6707 &	y &	1 &	255.3 &	0.7 &	0.0972 &	0.7 &	1.8 : &	$-$29.2 &	$-$0.017 &	Ole2004a  \\ 
22408$-$0333 &	KUI 114 &	HIP 111965 &	9.6707 &	I &	4 &	A291.7&	0.3 &	0.0508 &	0.5 &	0.8   &	$-$6.8 &	0.003 &	Sod1999  \\ 
22552$-$0459 &	BU 178 &	HIP 113184 &	9.6707 &	I &	3 &	322.2 &	0.7 &	0.6466 &	0.7 &	2.5   &	$-$1.1 &	$-$0.129 &	Baz1981b  \\ 
22553$-$4828 &	I 22AB &	HIP 113191 &	9.6679 &	y &	2 &	174.4 &	2.8 &	0.4450 &	0.7 &	0.9 :  &  &   &   \\ 
            &	       &	           &	9.6679 &	I &	2 &	174.8 &	0.2 &	0.4443 &	0.3 &	0.2    &  &   &   \\ 
22586$-$4531 &	HU 1335 &	HIP 113454 &	9.6679 &	I &	6 &	174.6 &	1.2 &	0.1605 &	2.3 &	0.6   &	$-$8.5 &	0.003 &	Hei1984a  \\ 
23052$-$0742 &	A 417 AB &	HIP 113996 &	9.6708 &	I &	2 &	A41.0&	0.1 &	0.2026 &	0.1 &	0.2   &	$-$0.7 &	$-$0.003 &	Hrt1996a  \\ 
            &	       &	           &	9.6708 &	y &	2 &	A40.9&	0.2 &	0.2031 &	0.1 &	0.5   &	$-$0.8 &	$-$0.002 &	Hrt1996a  \\ 
23099$-$2227 &	RST 3320 &	HIP 114375 &	9.6707 &	y &	2 &	135.1 &	0.2 &	0.1843 &	0.1 &	1.1   &	$-$43.2 &	$-$0.079 &	Sey2002  \\ 
            &	       &	           &	9.6708 &	I &	3 &	135.1 &	0.8 &	0.1851 &	0.9 &	2.0   &	$-$43.2 &	$-$0.078 &	Sey2002  \\ 
23191$-$1328 &	MCA 74 Aa,Ab &	HIP 115126 &	9.6708 &	I &	2 &	A201.5&	1.4 &	0.1834 &	0.6 &	2.9   &	$-$9.7 &	0.016 &	Msn1999c  \\ 
            &	       &	           &	9.6708 &	y &	1 &	A201.9&	1.3 &	0.1822 &	1.3 &	3.2 : &	$-$9.4 &	0.014 &	Msn1999c  \\ 
23227$-$1502 &	HU 295 &	HIP 115404 &	9.6707 &	I &	2 &	281.0 &	0.3 &	0.3206 &	0.2 &	1.1   &	$-$0.3 &	$-$0.010 &	Sey1999b  \\ 
            &	       &	           &	9.6707 &	y &	2 &	281.0 &	0.3 &	0.3203 &	0.3 &	1.4 : &	$-$0.3 &	$-$0.010 &	Sey1999b  \\ 
            &	       &	           &	9.6763 &	I &	13 &	281.1 &	0.8 &	0.3215 &	1.5 &	1.1   &	$-$0.3 &	$-$0.009 &	Sey1999b  \\ 
            &	       &	           &	9.6763 &	V &	2 &	281.0 &	0.7 &	0.3200 &	1.5 &	1.2 : &	$-$0.4 &	$-$0.010 &	Sey1999b  \\ 
23357$-$2729 &	SEE 492 &	HIP 116436 &	9.6708 &	I &	2 &	16.2 &	0.5 &	0.6355 &	0.2 &	1.6   &	2.9 &	0.053 &	Hei1984a  \\ 
23529$-$0309 &	FIN 359 &	HIP 117761 &	9.6708 &	I &	2 &	A355.7&	0.2 &	0.0551 &	0.1 &	0.3   &	5.3 &	0.006 &	Msn2010  \\ 
23586$-$1408 &	RST 4136 AB &	HIP 118205 &	9.6708 &	I &	2 &	A208.2&	0.4 &	0.1511 &	0.4 &	0.9 : &	$-$7.3 &0.035 &	Msn1999c  \\ 
23587$-$0333 & BU 730       &  HIP 118209  &    9.7553 &        I &     2 &     322.1 & 19.1&   0.7571 &        8.4 &   4.7   & $-$1.0 & $-$0.129 & Sey2002 \\
\enddata   
~  \\  
\vskip 0.1in                                                                                                                                                                    
* The complete list of references may be found at http://ad.usno.navy.mil/Webtextfiles/wdsnewref.txt .                                                                          
\end{deluxetable}                                                                                                                                                               




\end{document}